\documentclass[showpacs,prb,twocolumn,floatfix,superscriptaddress,amsmath]
{revtex4}
\usepackage{graphicx,amsmath}

\begin{document}

\title{Dynamic correlations in symmetric electron-electron and electron-hole
 bilayers} 

\author{R. K. Moudgil} \altaffiliation{On study leave from Department of
Physics, Kurukshetra University, Kurukshetra- 136 119, India.}
\affiliation{Dipartimento di Fisica Teorica, Universit\`a di Trieste, Strada
Costiera 11, I-34014 Trieste, Italy.}

\author{L. K. Saini} 
\affiliation{Department of Physics, Panjab University,  Chandigarh - 160014,
 India.}

\author{Gaetano Senatore} \email[]{senatore@ts.infn.it}
\affiliation{INFM -- DEMOCRITOS National Simulation Center}
\affiliation{Dipartimento di Fisica Teorica, Universit\`a di Trieste, Strada
Costiera 11, I-34014 Trieste, Italy.}

\date{\today}

\begin{abstract}
The ground-state behavior of the symmetric electron-electron and electron-hole
bilayers is studied by including dynamic correlation effects within the
quantum version of Singwi, Tosi, Land, and Sj\"{o}lander (qSTLS) theory.  The
static pair-correlation functions, the local-field correction factors, and the
ground-state energy are calculated over a wide range of carrier density and
layer spacing.  The possibility of a phase transition into a density-modulated
ground state is also investigated.  Results for both the electron-electron and
electron-hole bilayers are compared with those of recent diffusion Monte Carlo
(DMC) simulation studies.  We find that the qSTLS results differ markedly from
those of the conventional STLS approach and compare in the overall more
favorably with the DMC predictions.  An important result is that the qSTLS
theory signals a phase transition from the liquid to the coupled Wigner
crystal ground state, in both the electron-electron and electron-hole
bilayers, below a critical density and in the close proximity of layers
($d\alt r_sa_0^*$), in qualitative agreement with the findings of the DMC
simulations.
\end{abstract}


\maketitle

\section{Introduction}
In recent years, there has been considerable interest in the study of systems
composed of two or more (equispaced) electron layers.  The advances in the
nanoscale semiconductor fabrication technology (such as the molecular beam
epitaxy, the lithography techniques, etc.)  have made available these electron
systems in the coupled semiconductor quantum-well structures with a good
control on the electron number density and the interwell spacing. A variety of
new interesting phenomena have been observed, due entirely to the presence
of interlayer Coulomb interactions.  The stability of new fractional quantum
Hall states \cite{r1} and the discovery of an insulating Wigner crystal (WC)
phase \cite{r2}, in the bilayer electron system in the presence of a
perpendicular magnetic field, are some prominent examples. Theoretical
\cite{r3,r3a} as well as recent diffusion Monte Carlo (DMC) \cite{r4a,r4}
studies have predicted, even in the absence of magnetic field, the
stabilization of the WC phase in the electron bilayer in the close proximity
of layers $d\alt r_sa_0^*$ (see below for the definition of relevant
parameters) at sufficiently low electron density. More precisely, the critical
density for the phase transition is predicted to shift towards the higher
density side as compared to the corresponding value in the case of an isolated
electron layer.  In the following we shall be confining our discussion to the
double layer system.

Central in understanding the behavior of the layered electron systems are the
intra- and interlayer many-body correlation effects.  Since the early
random-phase approximation (RPA) study by Das Sarma and Madhukar \cite{r5},
there have been many investigations focusing on the role of many-body
correlations. The theoretical techniques used have mostly relied on the
extensions to two dimensions (2D) of existing theories.  Zhang and Tzoar
\cite{r6}, Neilson and co-workers \cite{r7}, and Zheng and MacDonald \cite{r8}
have used the mean-field approximation of Singwi, Tosi, Land, and
Sj\"{o}lander (STLS) \cite{r9} to study the effect of correlations on the
various ground-state properties of an electron bilayer system.  On the other
hand, Kalman and co-workers \cite{r10} have incorporated correlations by
satisfying the third-frequency sum rule of the density-density response
function to study the collective modes.  A general conclusion has emerged that
the interlayer interactions add further to the importance of many-body
correlations, which are already known to be very important in an isolated
electron layer \cite{rjon}.  Both in the STLS and Kalman's group approaches,
correlations enter in the theory in the form of a static local-field
correction (LFC). In STLS, the LFC's are obtained numerically in a
self-consistent way, while in the latter approach their calculation rely on
the knowledge of accurate pair-correlation functions.  The extent of validity
of the STLS or other theories in the present context can be tested by making a
direct comparison with the accurate DMC simulation results of Rapisarda and
Senatore \cite{r4a,r4}. However, no such comparison of the STLS results has been
made so far. A comparative study is available only in case of an isolated
electron layer \cite{r11}, where it has been found that the STLS theory,
although providing a significant improvement over the lower order random-phase
and Hubbard \cite{hubb} approximations, yet it fails to give an adequate
description of correlations beyond $r_s>3$.  In particular, it yields negative
value for the pair-correlation function (an unphysical result) at small
separation for $r_s>3$.  This failure of the STLS approach also appears in the
bilayer problem \cite{r8}.  As usual, here $r_s=1/(a_0^*\sqrt{n\pi})$ is the
dimensionless density parameter, with $a_0^*=\hbar ^2/(m_e^*e^2)$ the
effective Bohr atomic radius and $n$ the in-layer areal number
density. $m_e^*$ is the effective (band) mass of electron. We recall that
$r_s$ also provides a rough estimate of the in-layer coupling, as ratio of the
(independent particle) potential and kinetic energies of the system.  The
failure of the STLS theory at higher $r_s$ has been traced back to its basic
assumption of treating as static (i.e., time-independent) the electronic
exchange-correlation hole\cite{r12}.

In the two-layer system, the interlayer interactions will result in an
increase in Coulomb coupling as compared to that in an isolated layer at the
same density.  Therefore, we anticipate that the dynamics of correlations will
be even more crucial in the two-layer system, with respect to an isolated
layer.  This forms part of our motivation for the present work.  Here, we
intend to examine the ground-state behavior of the symmetric electron-electron
(e-e) bilayer by including dynamic electron correlation effects. To this end,
we make use of the dynamic or quantum version of the STLS approach
(qSTLS). The qSTLS theory embodies correlations beyond the conventional STLS
approach and as an important improvement, its LFC is frequency-dependent. It
should be pointed out here that the so-called qSTLS theory was originally
developed by Hasegawa and Shimizu \cite{r13} for the 3D electron system, and
that its predictions of many-body properties, as adjudged by comparison with
the MC results, are better irrespective of the dimensionality \cite{r11} and
the carrier statistics \cite{r14}.  In view of the very recent DMC study of
the symmetric electron-hole (e-h) bilayer by De Palo, Rapisarda, and Senatore
\cite{r15}, we also employ the qSTLS theory to examine the ground-state
properties of such system.  For both the e-e and e-h bilayers we present
results for the pair-correlation functions, the LFC factors, and the
ground-state energy over a wide range of density and layer spacing.  We also
look for signs indicating an instability of the homogeneous liquid with
respect to inhomogeneous phases of the charge-density-wave (CDW) and WC
type. Finally, we compare our results with those of the DMC studies.

The rest of the paper is organized as follows: In Sec. II, we present in brief
the qSTLS formalism for the double-layer system.  Results and discussion are
presented in Sec. III. In Section IV, we conclude the paper with a brief
summary.

\section{Theoretical formalism}
\subsection{Model}
We consider a double quantum-well structure with $d$ as the center-to-center
well separation. The carriers are electrons in one well and electrons or holes
in the other, respectively for the e-e and e-h bilayer.  The motion of
carriers is free along the $xy$-plane and under the action of a double well
potential profile in the $z$-direction. We assume that the wells are extremely
narrow and the potential barriers along the $z$-axis are high enough so that
the particles occupy only the lowest energy subband for the $z$ motion and
there is negligible overlap between the wave functions of particles in
the two wells.  The wells are assumed to be identical in each respect except
for the charge of carriers in the e-h bilayer. Further, the bilayer system is
assumed to be embedded in a uniform charge neutralizing background.  On
neglecting the effect of integrating over the finite extent of the particle
wave function in the $z$-direction, the Coulomb interaction potential among
the carriers is obtained as 
\begin{equation} 
V_{ll'}(q)=\alpha_{ll'} V(q)e^{-q|l-l'|d},
\label{e1}
\end{equation} 
with $l=1,2$ the layer index and $V(q)=2\pi e^2/(q\epsilon_0)$ the intralayer
interaction potential. Above $\epsilon_0$ is the background dielectric
constant and $\alpha_{ll'}=1$ and $(-1)^{|l-l'|}$, respectively, for the e-e
and e-h bilayers.

Apparently, the ground state of the above bilayer model will depend, apart
from $r_s$, on the interlayer spacing $d$.  It turns out convenient to
introduce an additional coupling parameter as the ratio of the typical
interlayer and in-layer Coulomb energies, namely, $\gamma=r_sa_0^*/d$. Thus,
at $T=0$, which is the case considered here, the bilayer model may be
completely specified by $r_s$ and $\gamma$ (or $d$).

\subsection{Density response function}
In the dielectric approach, the density-density linear response function
$\chi(q,\omega)$, which describes the response to an external potential
$V^{ext}(q,\omega)$ that couples to the particle density, plays the role of a
central quantity in determining the many-body properties of the system.  For
the bilayer, the linear response matrix $\chi_{ll'}(q,\omega)$ is formally
defined by 
\begin{equation} 
\delta\rho_l(q,\omega)=
\sum_{l'=1}^{2}\chi_{ll'}(q,\omega)V_{l'}^{ext}(q,\omega),\label{e2}
\end{equation} 
where $\delta\rho_l(q,\omega)$ represents the induced particle density in the
$l$th layer and $V^{ext}_{l}(q,\omega)$ the potential that couples to the
density in the layer $l$.  For completeness, we give in the following a very
brief account of the qSTLS formulation of the density response for the
two-layer system.

The equation of motion for the one-particle Wigner distribution function (WDF)
$f^{\sigma}_{l}({\bf r},{\bf p};t)$ (the superscript $\sigma$ is the spin
index) involves\cite{r11}, the {\sl unknown} two-particle WDF $f^{\sigma
\sigma '}_{ll'}({\bf r},{\bf p};{\bf r'},{\bf p'};t)$.  Progress can however
be made resorting to the STLS approximate decoupling ansatz \cite{r9}, which
in the two-layer system becomes
\begin{equation} 
f^{\sigma \sigma '}_{ll'}({\bf r},{\bf p};{\bf r'},{\bf p'};t)
=f^{\sigma}_{l}({\bf r},{\bf p};t)f^{\sigma '}_{l'}({\bf r'},{\bf p'};t)
g^{\sigma \sigma '}_{ll'}(|{\bf r}-{\bf r'}|),
\label{e3}
\end{equation} 
with $g^{\sigma \sigma '}_{ll'}(|{\bf r}-{\bf r'}|)$ the equilibrium static
pair-correlation function between carriers of spin $\sigma$ and $\sigma'$ in
the layers $l$ and $l'$.  Expressing the particle density in terms of the the
one-particle WDF\cite{r13}, one readily obtains the induced density
$\delta\rho_l(q,\omega )$ in the $l$th layer as 
\begin{equation}
\delta\rho_l(q,\omega)=\chi_l^0(q,\omega)\left
[V^{ext}_{l}(q,\omega)+V^{pol}_{l}(q,\omega)\right],
\label{e4}
\end{equation} 
with 
\begin{equation}
V^{pol}_{l}(q,\omega)=\sum_{l'=1}^{2}\delta\rho_{l'}(q,\omega)
V_{ll'}(q)[1-G_{ll'}(q,\omega)]
\label{e5}
\end{equation} 
the {\sl polarization} potential, $\chi_l^0(q,\omega)$ the density response
function of non-interacting electrons in layer $l$ (i.e., the Stern function
\cite{r16}) and
\begin{widetext}
\begin{equation} 
G_{ll'}(q,\omega)=-\frac{1}{n}\int\frac{d{\bf q'}}{(2\pi)^2}
\frac{\chi_l^0({\bf q},{\bf q'};\omega)V_{ll'}(q')}
{\chi_l^0(q,\omega)V_{ll'}(q)} [S_{ll'}(|{\bf q}-{\bf q'}|)-\delta_{ll'}],
\label{e6}
\end{equation}
\end{widetext}
the {\sl dynamic} LFC factor that accounts for correlation effects among
carriers in the layers $l$ and $l'$.  In Eq. (\ref{e6}), $S_{ll'}(q)$ is a
static structure factor and $\chi_l^0({\bf q},{\bf q'};\omega)$ the
inhomogeneous Stern function given by
\begin{equation} 
\chi_l^0({\bf q},{\bf q'};\omega)=-2\int\frac{d{\bf k}}
{(2\pi)^2}\frac{f_l^0({\bf k}+{\bf q'}/2)-f_l^0({\bf k}-{\bf q'}/2)}
{\omega-\hbar {\bf k}\cdot{\bf q}/m+\iota\eta},
\label{e7}
\end{equation}
where $f_l^0({\bf k})$ is the usual non-interacting Fermi-Dirac distribution
function and $\eta$ is a positive infinitesimal. For ${\bf q'}={\bf q}$,
$\chi_l^0({\bf q},{\bf q'};\omega)$ reduces to Stern function
$\chi_l^0(q,\omega)$.  Using Eqs. (\ref{e2}), (\ref{e4}), and (\ref{e5}) the
elements of the inverse of the linear response matrix are readily obtained as
\begin{equation} 
{\chi^{-1}}_{ll'}(q,\omega) = \frac{\delta_{ll'}}{\chi_l^{0}(q,\omega)}
-V_{ll'}(q)[1-G_{ll'}(q,\omega)].
\label{e8}
\end{equation}

The fluctuation-dissipation theorem, which relates the static structure
factors with the imaginary part of the linear response functions as
\begin{equation} 
S_{ll'}(q)=-\frac{\hbar}{n\pi}\int_{0}^{\infty} d\omega
Im\,\chi_{ll'}(q,\omega),
\label{e9}
\end{equation} 
closes the qSTLS set of equations for the density response matrix.  Evidently,
the response function calculation has to be carried out numerically in a
self-consistent way. In view of the symmetry of the bilayer, we will have
$A_{11}=A_{22}$ and $A_{12}=A_{21}$, where $A$ refers to the general layer
property.

\subsection{Pair-correlation function and ground-state energy}
The pair-correlation function $g_{ll'}(r)$ can be obtained directly from the
inverse Fourier transform of the static structure factor as 
\begin{equation}
g_{ll'}(r)=1+\frac{1}{n}\int\frac{d{\bf q}}{(2\pi)^2}e^{\iota {\bf q}.{\bf r}}
[S_{ll'}(q)-\delta_{ll'}].
\label{e10}
\end{equation} 
The ground-state energy $E_{gs}$ (defined here as per particle) is determined
by a straightforward extension of the ground-state energy theorem \cite{mahan}
to the two-layer system as 
\begin{equation}
E_{gs}=E_0+\int_{0}^{e^2}\frac{d\lambda}{\lambda}E^{int}(\lambda),
\label{e11}
\end{equation}
where $E_0=p_F^2/(4m^*_e)$ is the kinetic energy per particle of the
non-interacting system ($p_F=\hbar q_F$, is the Fermi momentum), $\lambda$ is
the strength of Coulomb interaction, and $E^{int}(\lambda)$ is the interaction
energy per particle given by 
\begin{equation}
E^{int}(\lambda)=\frac{1}{4}\sum_{l,l'}^{2}\int\frac{d{\bf q}}{(2\pi)^2}
\lambda\, V_{ll'}(q)[S_{ll'}(q; \lambda)-\delta_{ll'}].
\label{e12}
\end{equation} 
In the next section, we present results for the ground-state properties of the
 e-h and e-e bilayers.

\section{Results and discussion} 
\subsection{Pair-correlation functions}
Equations (\ref{e6}), (\ref{e8}), and (\ref{e9}) are solved numerically in a
self-consistent way for $S_{ll'}(q)$. The $\omega$-integration in the
computation of $S_{ll'}(q)$ (Eq. (\ref{e9})) is performed along the imaginary
$\omega$-axis in order to avoid the problem of dealing with the plasmon poles
which appear on the real $\omega$-axis (see de Freitas et al \cite{r17} and
Ref. \cite{r11}).  We accepted the solution when the convergence in
$S_{ll'}(q)$ at each $q$ in the grid of $q$-points was better than $0.001\%$.
It is important to point out here that in all our calculations for the e-h
bilayer we have taken $m_h^*/m_e^*=1$ ( $m_h^*$ is the effective mass of
hole).

\begin{figure*}
\includegraphics[width=75mm,height=70mm]{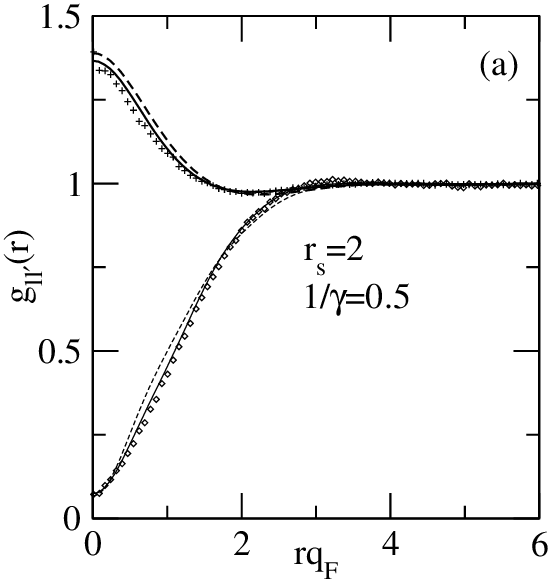}\vspace{5mm}
\includegraphics[width=75mm,height=70mm]{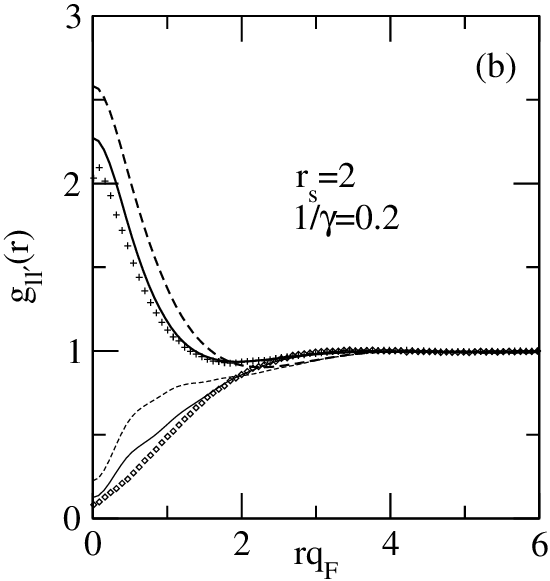}\vspace{5mm}
\includegraphics[width=75mm,height=70mm]{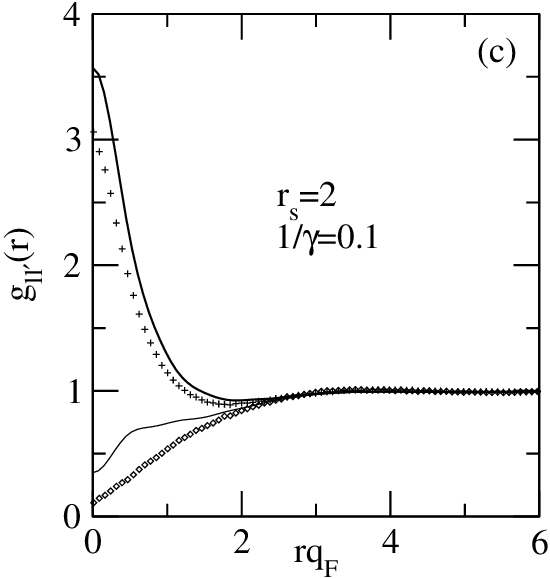}\vspace{5mm}
\includegraphics[width=75mm,height=70mm]{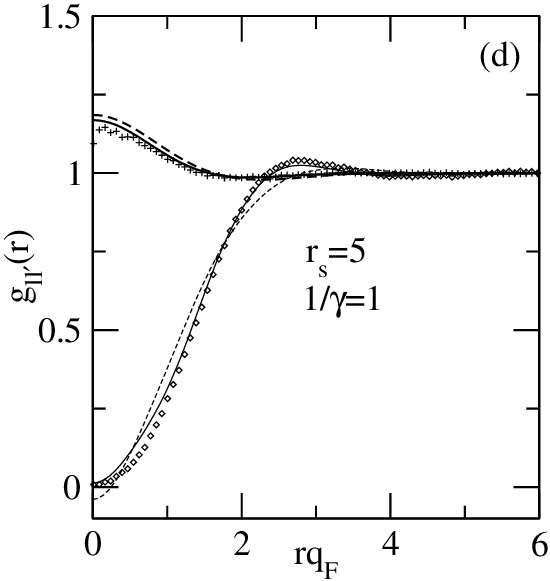}\vspace{5mm}
\includegraphics[width=75mm,height=70mm]{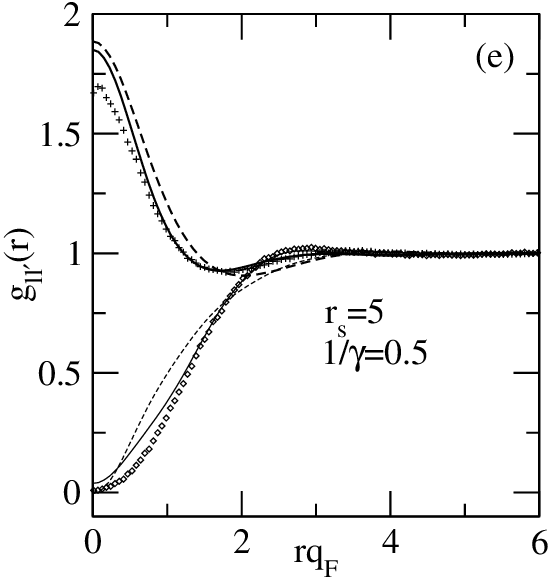}\vspace{5mm}
\includegraphics[width=75mm,height=70mm]{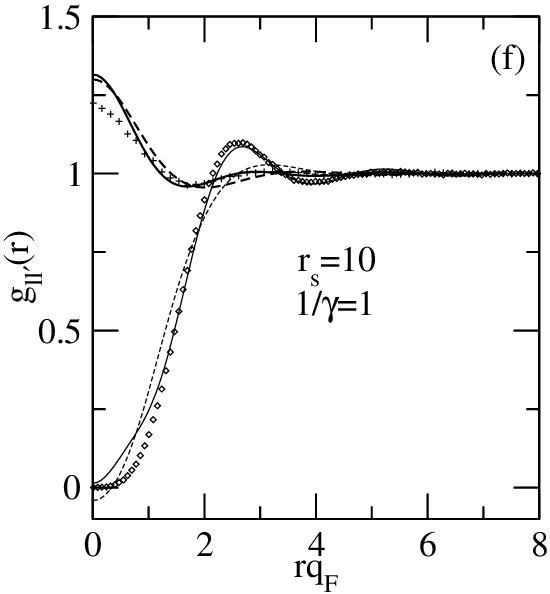}
\end{figure*}

\begin{figure}
\includegraphics[width=75mm,height=70mm]{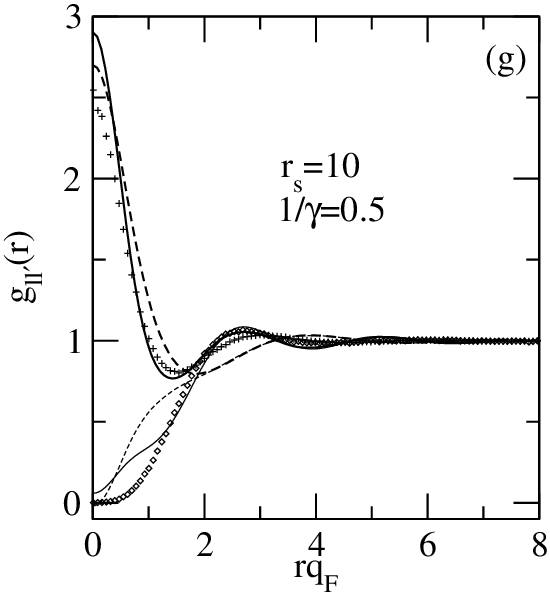}\vspace{5mm}
\includegraphics[width=75mm,height=70mm]{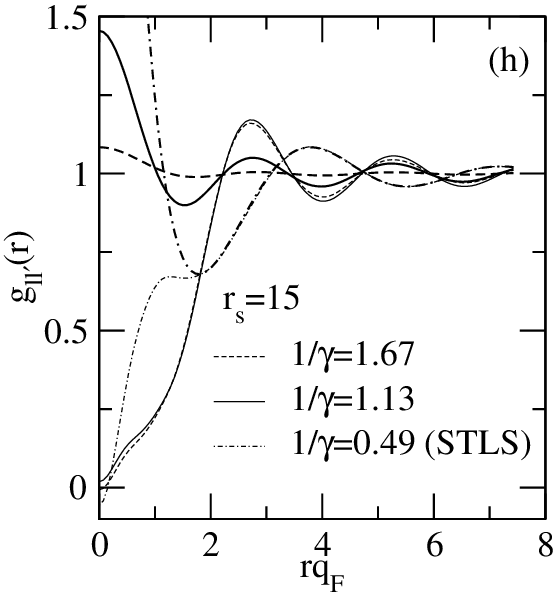}
\caption{\label{fig1}(a)-(g) Pair-correlation functions $g_{11}(r)$ (thin
lines) and $g_{12}(r)$ (thick lines) for the electron-hole bilayer at
different values of $r_s$ and $1/\gamma=d/r_sa_0^*$. Solid and dashed lines
are, respectively, the qSTLS and STLS results; the prediction of DMC
simulations\cite{r15,private_eh} for $g_{11}(r)$ ($\diamond$) and $g_{12}(r)$
($+$) are also shown.  (h) $g_{11}(r)$ (thin lines) and $g_{12}(r)$ (thick
lines) at indicated $r_s$ and $1/\gamma$ values, according to qSTLS; STLS
results are shown by dash-dot lines.}
\end{figure}

\begin{figure}
\includegraphics[width=75mm,height=70mm]{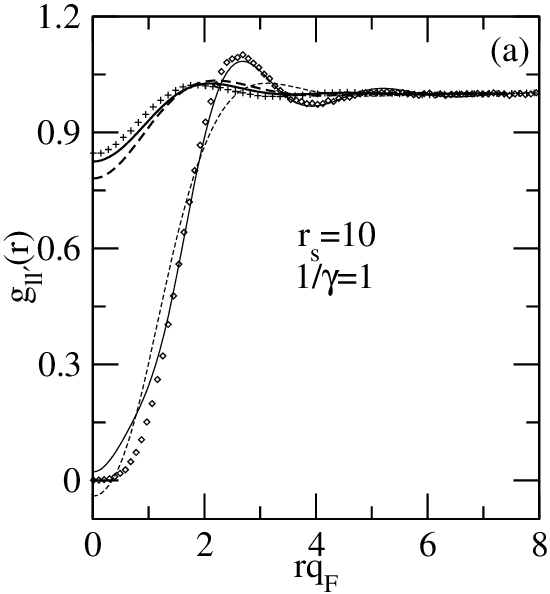}\vspace{5mm}
\includegraphics[width=75mm,height=70mm]{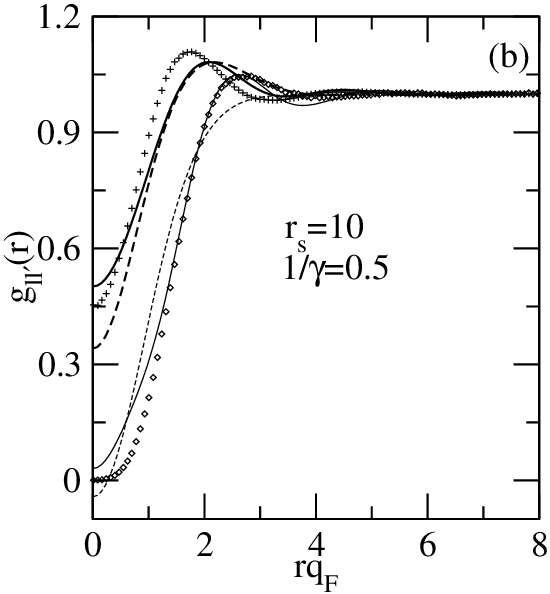}
\caption{\label{fig2}(a)-(b) Pair-correlation functions $g_{11}(r)$ and
$g_{12}(r)$ for the electron-electron bilayer; Curves are labeled as in
Fig. \ref{fig1}. The results of DMC simulations are from
Refs. \onlinecite{r4a,private_ee}.}
\end{figure}

Figures \ref{fig1} ((a)-(g)) and \ref{fig2} ((a)-(b)) show results for the
intra- and interlayer pair-correlation functions, $g_{11}(r)$ and $g_{12}(r)$,
for the e-h and e-e bilayers, respectively, at $r_s$ ($\leq 10$) and $d$
values where the DMC results are available for comparison. In order to have a
close comparison between the qSTLS and STLS results, the STLS curves are also
depicted in the same figures.  We first discuss the results for the e-h
bilayer: Looking at Fig. \ref{fig1} ((a)-(g)), we infer immediately that the
pair-correlation functions in qSTLS are in overall better agreement with the
DMC data than those in STLS. Among the notable features, the qSTLS theory, as
a marked improvement over STLS, accounts fairly well for the oscillatory
structure that develops in the DMC $g_{11}(r)$ and $g_{12}(r)$ with increasing
$r_s$, both in terms of amplitude and period.  The qSTLS $g_{11}(r)$ satisfies
the positive definiteness criteria of probability for $r_s$ up to $10$,
whereas the STLS $g_{11}(r)$ becomes slightly negative at small $r$ for
$r_s>3$.  We also notice that, common at each $r_s$, the quality of agreement
between theory and DMC data somewhat diminishes for increasing values of
$\gamma$. Moreover, quite similar is the trend of agreement with increasing
$r_s$ at a given value of $\gamma$; for instance, compare $g_{11}(r)$ and
$g_{12}(r)$ at $1/\gamma=0.5$ and $r_s=2,5,10$.  This shortcoming of the qSTLS
theory seems to stem from the fact that the frequency-dependence in its LFC's
represents a quantum-mechanical correction to the conventional STLS theory,
while the dynamics of spatial correlations among carriers, which is expected
to become vital at higher values of Coulomb coupling, is still missing. It can
in fact be shown \cite{r11} that the qSTLS LFC's reduce formally to the
frequency-independent STLS LFC's in the limit $\hbar \rightarrow 0$.

Since decreasing $d$ at a given $r_s$ or increasing $r_s$ at a given $d$
result in an increase in the Coulomb coupling among carriers in the bilayer,
the qSTLS assumption of static spatial correlations (i.e., the assumption of
using static pair-correlation function in Eq. (\ref{e3})) is expected to
become relatively less reliable at larger values of $\gamma$ or $r_s$.
Further, we find that it becomes almost impossible to obtain the
self-consistent solution in both the qSTLS and STLS above a critical value of
$\gamma$ ($r_s$) at a given $r_s$ ($\gamma$). The critical parameters are of
course different in the two approaches.  This is the reason that the STLS
curves are absent at $r_s=2$ for $1/\gamma=0.1$ in Fig. \ref{fig1} (c).  The
difficulty in obtaining the self-consistent solution, as we will see in detail
in a subsequent section, is related to the instability of the system against
transition to a density-modulated ground state.

\begin{table}
\caption {\label{tab1}The ground-state energy $E_{gs}$ per particle (in units
of effective Rydberg) at different $r_s$ and $1/\gamma$ values
($\gamma=r_sa_0^*/d$) for the e-h bilayer, according to qSTLS and STLS. DMC
results are from Ref. \onlinecite{r15}.}
\begin{ruledtabular}
\begin{tabular}{rcccc} 
$r_s$ & $1/\gamma$ & qSTLS & DMC & STLS \\ 
\hline  
2 & 0.1 & -0.6519 & -0.6947 & - \\ 
2 & 0.2 & -0.5831 & -0.6116 & -0.5757 \\ 
2 & 0.5 & -0.5412 & -0.5405 & -0.5307 \\ 
5 & 0.5 & -0.3057 & -0.3125 & -0.3010 \\ 
5 & 1.0 & -0.3015 & -0.3009 & -0.2970 \\ 
10 & 0.5 & -0.1724 & -0.1801 & -0.1706 \\
10 & 1.0 & -0.1697 & -0.1715 & -0.1682 \\ 
10 & 1.5 & -0.1695 & -0.1703 & -0.1680  
\end{tabular}
\end{ruledtabular}
\end{table}
\begin{table}
\caption {\label{tab2} The ground-state energy $E_{gs}$ per particle (in units
of effective Rydberg) at $r_s=10$ for different $1/\gamma$ values
($\gamma=r_sa_0^*/d$) for the e-e bilayer, according to qSTLS and STLS. DMC
results are from Ref. \onlinecite{r4a}.}
\begin{ruledtabular}
\begin{tabular}{ccccc} 
$r_s$ & $1/\gamma$ & qSTLS & DMC & STLS \\ 
\hline 
10 & 0.5 & -0.1707 & -0.1781 & -0.1699 \\ 
10 & 1.0 & -0.1685 & -0.1713 & -0.1677 \\ 
10 & 1.5 & -0.1683 & -0.1705 & -0.1675 
\end{tabular}
\end{ruledtabular}
\end{table}
For the e-e bilayer, the DMC pair-correlation functions are at present
available only in the strong coupling region $r_s\geq 10$.  Figure \ref{fig2}
((a)-(b)) present a comparison of our results at $r_s=10$ for $1/\gamma=1$ and
$0.5$.  We notice that the qSTLS provides a reasonable estimate for
$g_{11}(r)$, but it fails ( together with STLS) to give a satisfactory
description of $g_{12}(r)$ at $1/\gamma=0.5$. This points once again to the
importance of dynamics of spatial correlations among electrons in the e-e
bilayer. The improvement over the STLS predictions is again quite noticeable.

\subsection{Ground-state energy}   

The self-consistently obtained static structure factors $S_{11}(q;\lambda)$
and $S_{12}(q;\lambda)$ are used in Eq. (\ref{e12}) to calculate
$E^{int}(\lambda)$ as a function of $\lambda$. The ground-state energy is then
determined by performing the coupling-constant ($\lambda$) integration in
Eq. (\ref{e11}). Results for the ground-state energy per particle for the e-h
and e-e bilayers are given, respectively, in Tables \ref{tab1} and \ref{tab2}
at $r_s$ ($\leq 10$) and $\gamma$ values where DMC data are available for
comparison. The STLS results are also reported. Apparently, the qSTLS results
compare more favorably with the DMC data. There is an increase in error with
respect to the DMC data with increasing $r_s$ at a given $\gamma$ and with
increasing $\gamma$ at a given $r_s$, which obviously is the reflection of the
behavior of the pair-correlation functions under these conditions.

\subsection{Density-modulated ground states} 
The DMC studies have predicted that both the e-e and e-h bilayers will favor
energetically the WC ground state above a critical value $r_s^c$ of the
in-layer coupling, at given $d$.  We calculate here the static ($\omega =0$)
generalized susceptibility (i.e., density-density response) in the liquid
phase to find out any evidence that the qSTLS theory might provide for the
transition to a WC ground state.  If such a transition does occur, it may
appear in the static susceptibility as a divergence at the reciprocal lattice
vector (RLV) of the WC lattice.  Diagonalizing the density response matrix
(\ref{e8}), the static susceptibility is obtained as
\begin{widetext} 
\begin{equation}
\chi_{\pm}(q,0)=\frac{\chi_1^0(q,0)}{1-\chi_1^0(q,0)
\left[V_{11}(q)(1-G_{11}(q,0))\pm V_{12}(q)(1-G_{12}(q,0))\right]}.
\label{e13} 
\end{equation} 
\end{widetext} 
The $+$ and $-$ signs correspond, respectively, to the in-phase and
out-of-phase ($\pi$) modes of density modulations $\delta \rho (q,0)$ in the
two layers.  An inspection of Eq. (\ref{e13}) makes it clear that
$\chi_{\pm}(q,0)$ can exhibit divergence at some $q$ value only if the
quantity within the square brackets in its denominator becomes sufficiently
negative (on account of the negative sign of $\chi_1^0(q,0)$) . In an isolated
layer (i.e., when $V_{12}(q)=0$) this can happen only if $G_{11}(q,0)$ has
values exceeding unity. However, in a bilayer \cite{r3,r7}, the interlayer
interaction term can cause a divergence even if $G_{11}(q,0)$ has values below
unity. Apparently, it is the in-phase component of susceptibility that can
have divergence in the e-h bilayer, while it is the out-of-phase component in
the e-e bilayer. Further, as the e-h and e-e correlations are of opposite
nature (i.e., $G_{12}(q,0)$ is negative in the e-h bilayer, while it is
positive in the e-e bilayer), they will act, respectively, to support and
oppose the formation of density-modulated phase (if any) in the e-h and e-e
bilayers.
 
The static LFC factors, $G_{11}(q,0)$ and $G_{12}(q,0)$, required in the
calculation of the susceptibility are determined by using the
self-consistently obtained static structure factors in Eq. (\ref{e6}). We find
quite generally that $\chi_{+}(q,0)$ ($\chi_{-}(q,0)$) exhibits for the e-h
bilayer (e-e bilayer) a strong peak-structure at a finite wavevector value in
the close proximity of two layers ($d\alt r_sa_0^*$).  A critical layer
separation $d_c$ (critical $r_s^c$) at a given $r_s$ ($d$) is encountered
below (above) which it becomes almost impossible to obtain the self-consistent
solution. Tracing carefully the different steps involved in the solution of
the qSTLS equations, the difficulty in obtaining the convergent solution is
found to be related directly to the emergence of the strong peak structure in
$\chi_{\pm}(q,0)$.  For $d<d_c$ at a given $r_s$ or for $r_s>r_s^c$ at given
$d$, a numerical instability (singularity) appears in $S_{ll'}(q)$ during the
iterative calculation, at a $q$ coinciding exactly with the peak position in
$\chi_{\pm}(q,0)$, while calculating $S_{ll'}(q)$ from $\chi_{ll'}(q,\omega)$
in Eq. (\ref{e9}).  We find it extremely difficult to handle the instability
in the self-consistent calculation and therefore, are unable\cite{note} to
find the convergent $S_{ll'}(q)$ below (above) a critical value of $d$
($r_s$). This forbids us, in turn, to calculate $\chi_{\pm}(q,0)$ below
(above) a critical value of $d$ ($r_s$) at a given $r_s$ ($d$).  On the other
hand, the emergence of a strong peak in $\chi_{\pm}(q,0)$ at a finite
wavevector value followed by the numerical instability at the same wavevector
during the self-consistent calculation of the density response functions could
be interpreted in our theory as an indication for the onset of a phase
transition from the liquid to the density-modulated ground state.

\subsubsection{Electron-hole bilayer}
\begin{figure}
\includegraphics[width=77mm,height=67mm]{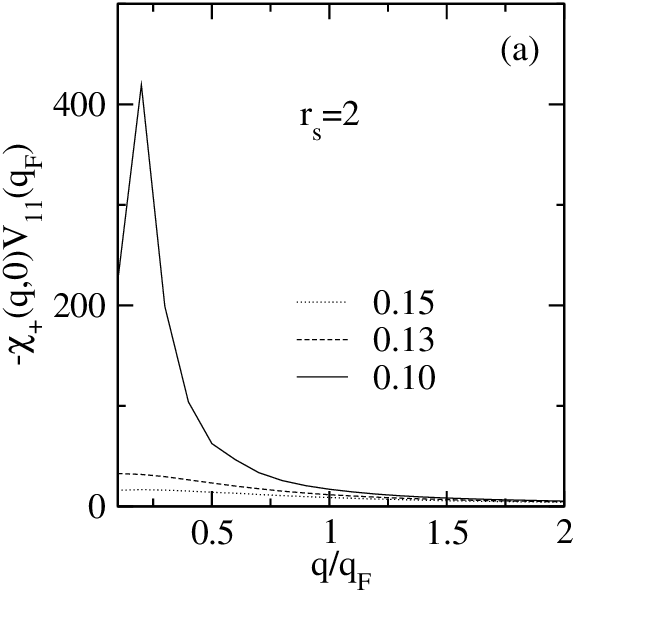}\vspace{5mm}
\includegraphics[width=77mm,height=67mm]{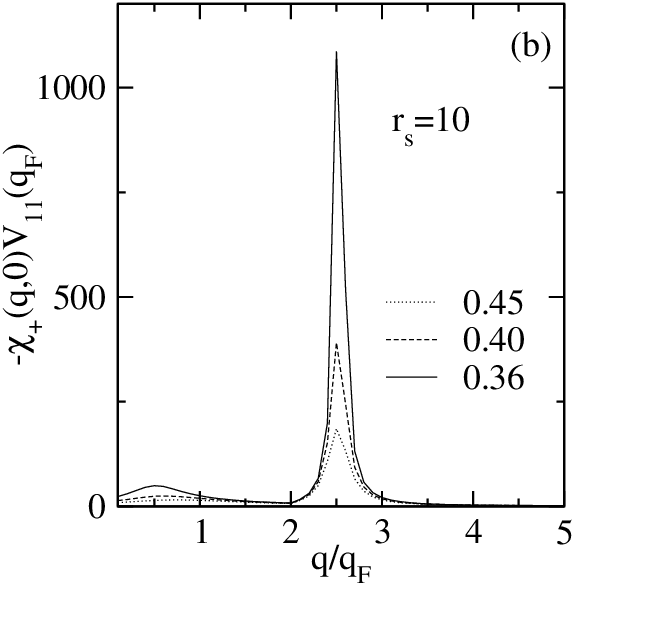}\vspace{5mm}
\includegraphics[width=77mm,height=67mm]{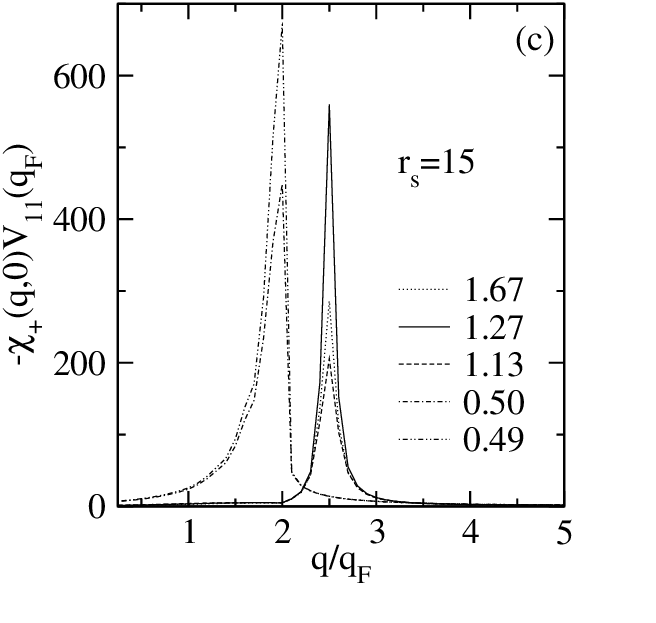}
\caption{\label{fig3}(a)-(c) In-phase component of the static density
susceptibility $\chi_{+}(q,0)$ for the electron-hole bilayer at different
values of $r_s$ and $1/\gamma=d/r_sa_0^*$, according to qSTLS.  Legends
indicate the values of $1/\gamma$; STLS results are shown for comparison by
dash-dot lines in (c) at $r_s=15$ and indicated $1/\gamma$ values.}
\end{figure}
Results for $\chi_{+}(q,0)$ at some selected values of $r_s$ and $d$ are shown
in Fig. \ref{fig3} ((a)-(c)).  For $r_s<5$, $\chi_{+}(q,0)$ has a single
strong peak at small $q$; for instance, at $r_s=2$ (Fig. \ref{fig3}(a)) the
peak is positioned at $q/q_F\approx 0.2$ and $d_c/(r_sa_0^*)\approx 0.095$.
For $r_s\geq 5$, however, a second peak starts developing in $\chi_{+}(q,0)$
at $q/q_F\approx 2.5$, with its strength relative to that of the peak at small
$q/q_F$ growing continuously as a function of increasing $r_s$.  At $r_s=10$
(Fig. \ref{fig3}(b)), the peak at $q/q_F\approx 2.5$ eventually dominates the
small-$q$ peak and $d_c/(r_sa_0^*)\approx 0.36$.  With further increase of
$r_s$, the small-$q$ peak disappears completely as shown in Fig. \ref{fig3}(c)
at $r_s=15$; $d_c/(r_sa_0^*)\approx 1.10$.  The position of the peak located
at $q/q_F\approx 2.5$ matches quite closely with the RLV for a triangular
Wigner lattice ($\approx 2.7$) and therefore, we speculate that this peak
signals a transition to the coupled in-phase WC ground state in the e-h
bilayer.  Our speculation draws further support from the fact that near the
transition point $g_{11}(r)$ and $g_{12}(r)$ exhibit strong in-phase
oscillations typical of an ordered phase, and this feature of $g(r)$ is
illustrated in Fig. \ref{fig1}(h) at $r_s=15$. On the other hand, the
small-$q$ peak, whose position varies with $r_s$, indicates the instability of
the e-h liquid against a CDW ground state.  Thus, there seems to occur a
crossover from a CDW to a WC ground state at a critical density $r_s^c\approx
10$.  The qSTLS indication of a transition to a WC state is in agreement
with the findings of the DMC study \cite{r15}. However, the critical $r_s$ for
the onset of the phase transition is underestimated by a factor of about
$2$. The qSTLS theory signals a Wigner crystallization also for an isolated
electron layer, at $r_s^c\approx 17$, again with an overestimate of the
critical coupling by a factor of about 2, as compared with the available DMC
predictions\cite{r4a,r4,r18,r18a}. Nevertheless, an important result to note
is that the electron-hole correlations in the e-h bilayer act to lower the
critical $r_s$ value, as compared to that in an isolated layer, by a factor of
about $1.7$, which matches closely with the DMC prediction.
 
Furthermore, the DMC study\cite{r15} has found that the WC ground state (also,
the liquid state when $r_s<r_s^c$) becomes energetically unstable against
transition to an excitonic ground state as the layers of electrons and holes
are brought close to each other ($d\alt r_sa_0^*$). There is some indication
in our results, in terms of the steady buildup of $g_{12}(r=0)$ in the close
vicinity of layers, for the formation of excitons \cite{exciton}, but there is
no apparent way in our theory to directly detect the transition to the
excitonic ground state.

It is appropriate at this point to draw a comparison of our results with the
previous work on the e-h bilayer by Liu et al \cite{r7} and Szymanski et al
\cite{r7}.  Liu et al treated the correlations within the STLS approach and
found at all $r_s$, in the close vicinity of layers, always an instability
towards the CDW ground state.  The STLS $\chi_{+}(q,0)$ is plotted for
comparison at $r_s=15$ in Fig. \ref{fig3}(c); the CDW wavevector $q/q_F$ is
$\approx 2$ and $d_c/(r_sa_0^*)\approx 0.48$.  Also shown for comparison in
Fig. \ref{fig1}(h) are the STLS $g_{11}(r)$ and $g_{12}(r)$ near the CDW
instability (at $d/(r_sa_0^*)=0.49$) at $r_s=15$.  On the other hand,
Szymanski et al employed the STLS local-fields to include the correlation
effects, but instead of carrying out the fully self-consistent STLS
calculation, they fixed the intralayer LFC $G_{11}(q)$ through the MC
structure factor for an isolated layer \cite{r18}, while the interlayer LFC
$G_{12}(q)$ was determined self-consistently by keeping $G_{11}(q)$ as a fixed
input.  Following this mixed STLS procedure, they predicted the presence of
both the CDW and WC instabilities, and also a crossover from the CDW state to
the WC state at $r_s\approx 15$.  Thus, there is a qualitative similarity
between the qSTLS results and the findings of Szymanski et al.

\begin{figure}
\includegraphics[width=75mm,height=70mm]{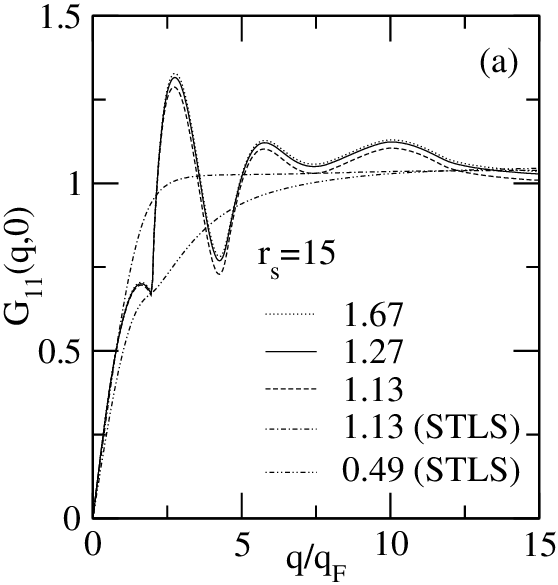}\vspace{5mm}
\includegraphics[width=75mm,height=70mm]{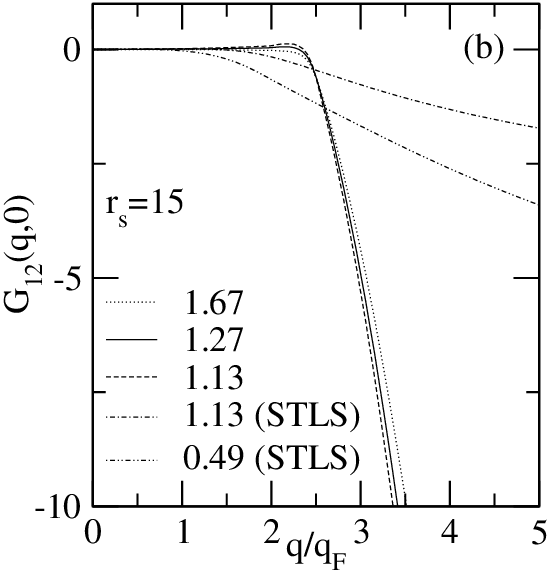}
\caption{\label{fig4}Intralayer (in panel (a)) and interlayer (in panel (b))
static local-field corrections $G_{11}(q,0)$ and $G_{12}(q,0)$ for the
electron-hole bilayer at $r_s=15$ and different layer spacings, according to
qSTLS. Legends indicate the values of $1/\gamma=d/r_sa_0^*$; STLS local-fields
are also shown.}
\end{figure}

It is gratifying to note that the qSTLS theory seems to capture at least
qualitatively the WC transition in the e-h bilayer.  The breakdown of the STLS
approach even at the qualitative level seems to have roots in its assumption
of treating correlations through the frequency-independent (i.e., static)
LFC's. To elucidate this viewpoint, the static ($\omega=0$) qSTLS LFC's and
the STLS LFC's, which in fact determine completely the behavior of
$\chi_{+}(q,0)$, are compared in Fig. \ref{fig4} at $r_s=15$ for different
$d's$. The LFC's differ markedly in the two approaches.  In particular, we
note that $G_{11}(q,0)$, in contrast with the STLS $G_{11}(q)$, exhibits an
oscillatory behavior, with pronounced maximum at $q/q_F\approx 3$ and with its
values lying well above unity in the relevant wavevector region of
$2<q/q_F<3$, before saturating to its limiting value of $(1-g_{11}(0))$. Also,
$G_{12}(q,0)$ is rapidly varying in the same $q-$region.  Consequently, the
qSTLS effective static intralayer interaction, namely
$[V_{11}(q)(1-G_{11}(q,0))]$, becomes attractive in this $q$-region, and this
effect of $G_{11}(q,0)$ in combination with the attractive interlayer e-h
correlations gives rise to a strong peak in $\chi_{+}(q,0)$ at $q/q_F\approx
2.5$. The static LFC factor exhibits a similar oscillatory behavior in an
isolated electron layer. Though the peaked-structure in the qSTLS static LFC
factor seems to enable the qSTLS theory to capture qualitatively the WC
instability, we have to mention that at least for an isolated layer such an
oscillatory structure is exaggerated as compared to that found in QMC
simulations\cite{r18b}.

\begin{figure}
\includegraphics[width=75mm,height=70mm]{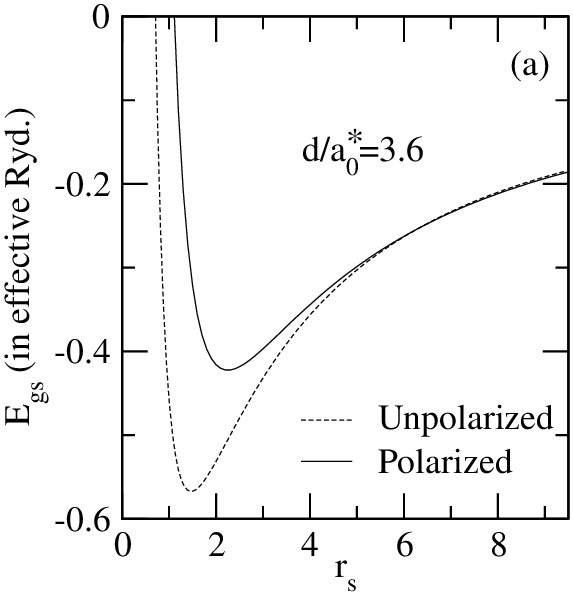}\vspace{5mm}
\includegraphics[width=75mm,height=70mm]{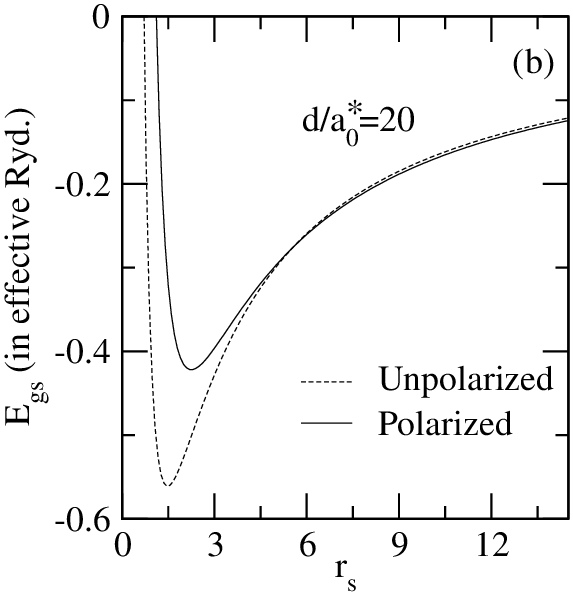}
\caption{\label{fig5}(a)-(b) Comparison of the ground-state energy per
particle $E_{gs}$ between the unpolarized and polarized phases of the
electron-hole bilayer at $d/a_0^*=3.6$ (in panel (a)) and $20$ (in panel (b)),
according to qSTLS.}
\end{figure}  
\begin{figure}
\includegraphics[width=75mm,height=70mm]{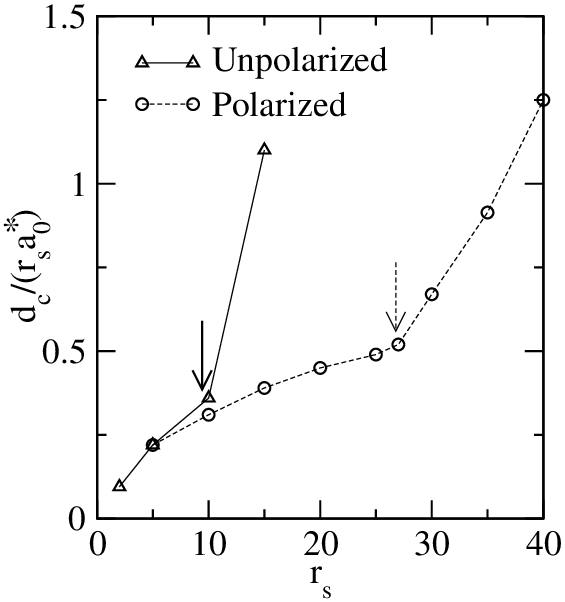}
\caption{\label{fig6}Critical layer spacing $d_c/(r_sa_0^*)$ (at a given
$r_s$) for the transition from the liquid to a density-modulated phase as a
function of $r_s$ in the unpolarized (triangles $\triangle$) and polarized
(circles $\circ$) phases of the electron-hole bilayer, according to qSTLS. For
each case the arrows show the critical density where crossover from the
charge-density-wave instability to the Wigner crystal instability occurs; The
lines are just a guide for the eye.}
\end{figure}

We also examine the stability of the ground state of the e-h bilayer as a
function of spin-polarization. For simplicity reasons, only two states of
spin-polarization, namely the fully polarized and unpolarized (which otherwise
is the case throughout the paper), are considered.  A comparison between the
ground-state energies of the e-h bilayer in the two states of
spin-polarization (Fig. \ref{fig5}) reveals that a polarization transition
occurs at a critical density in the liquid phase from the unpolarized to the
polarized state before the unpolarized liquid could actually make transition
to the seemingly WC phase.  The critical density for polarization transition
depends only weakly on $d$; for instance the critical $r_s$ decreases from $6$
to $5.5$ as $d/a_0^*$ increases from $3.6$ to $20$. We emphasize here that
$d/a_0^*=3.6$ is the critical value of layer spacing at $r_s=10$ for the WC
instability.  In this perspective, it becomes interesting to investigate the
ground state of the polarized e-h bilayer. We find after analyzing the static
susceptibility results that the qualitative behavior of the e-h ground state
does not depend upon the spin-polarization.  However, the crossover from the
CDW phase to the WC phase now occurs at $r_s\approx 27$. The dependence of the
point of instability on the spin-polarization is depicted in
Fig. \ref{fig6}. Evidently, the critical spacing $d_c$ in the polarized phase
lies always below to that in the unpolarized phase.

\subsubsection{Electron-electron bilayer}
In contrast with the e-h bilayer, the interlayer correlations in the e-e
bilayer tend to oppose the transition to the density-modulated phase.  But,
we find that this tendency of the interlayer correlations depends crucially on
the rate of their growth with  decreasing $d$, and that this rate is not
strong enough to preclude transition to the density-modulated phase.  We find
indication for both the CDW and WC instabilities.

\begin{figure}
\includegraphics[width=75mm,height=70mm]{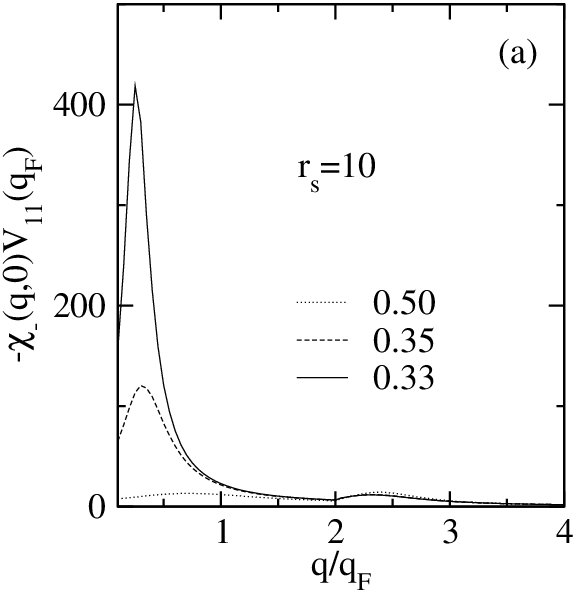}\vspace{5mm}
\includegraphics[width=75mm,height=70mm]{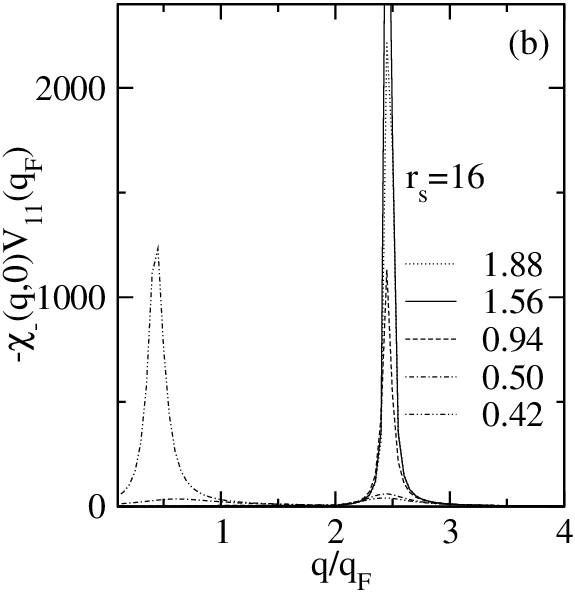}
\caption{\label{fig7}(a)-(b) Out-of-phase component of the static density
susceptibility $\chi_{-}(q,0)$ for the electron-electron bilayer at $r_s=10$
and $16$ and different values of $1/\gamma=d/r_sa_0^*$, according to
qSTLS; legends give the $1/\gamma$ values.}
\end{figure}

\begin{figure}
\includegraphics[width=75mm,height=70mm]{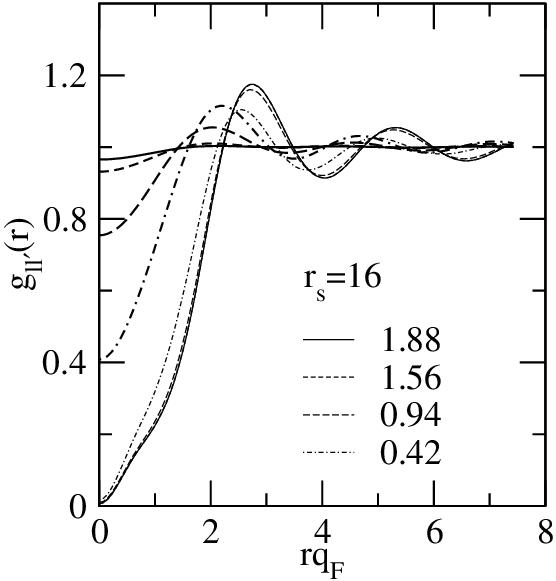}
\caption{\label{fig8}Pair-correlation functions $g_{11}(r)$ (thin lines) and
$g_{12}(r)$ (thick lines) in the electron-electron bilayer at $r_s=16$ and
different values of $1/\gamma=d/r_sa_0^*$, according to qSTLS. Legends give
the $1/\gamma$ values. Note that the $g_{11}(r)$ at $1/\gamma=1.88$ and at
$1/\gamma=1.56$ are practically indistinguishable.}
\end{figure}

The $\chi_{-}(q,0)$ results are shown in Fig. \ref{fig7} at $r_s=10$ and $16$
for some selected $d$'s.  The CDW instability now completely dominates for
$r_s$ up to 15. But, we notice an interesting behavior of $\chi_{-}(q,0)$ at
$r_s=16$. $\chi_{-}(q,0)$ exhibits a single strong peak at $q/q_F\approx 2.5$
when the layers are widely separated. The peak-height initially increases with
decrease in $d$, nearly diverges at $d/(r_sa_0^*)\approx 1.56$, and then
decreases monotonically with further decrease in $d$. There starts developing,
however, a second peak in $\chi_{-}(q,0)$ at $q/q_F\approx 0.5$ for
$d/(r_sa_0^*) \leq 0.5$, and this peak eventually dominates at $d/(r_sa_0^*)
\approx 0.45$ and then appears to diverge for $d/(r_sa_0^*) <0.42$.  This
might possibly imply that there is a crossover to the WC ground state at
$r_s\approx 16$, with the WC state however remaining stable only over a
certain range of layer spacings, with a transition back to the CDW-like phase
when the distance is further diminished. The pair-correlation functions in the
relevant range of $d$ values are shown in Fig. \ref{fig8}. Evidently, there is
a continuous decline in the amplitude of oscillation in $g_{11}(r)$ with
decreasing $d/(r_sa_0^*)$ below $1.56$ pointing to the obvious screening of
in-layer correlations by the interlayer interactions.  The present suggestion
of the stability of the WC ground state in the e-e bilayer only at
intermediate distances is compatible with the findings of DMC
calculations\cite{r4a,r4}, which for $r_s$ not too large predict the WC phase
to remain stable only at intermediate layer spacing, with the liquid phase
becoming stable again at smaller spacing.

\begin{figure}
\includegraphics[width=75mm,height=70mm]{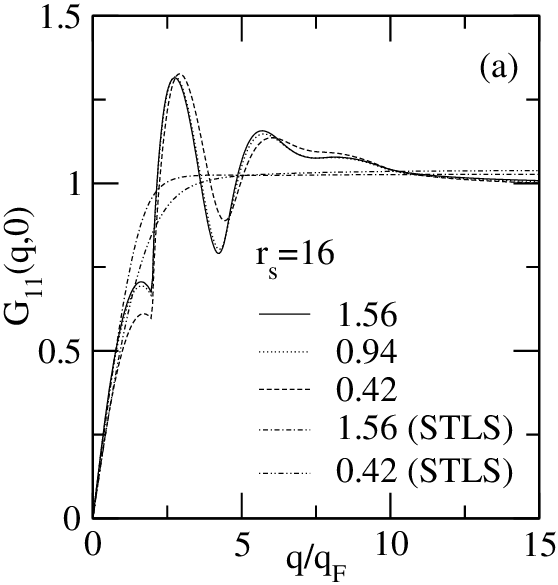}\vspace{5mm}
\includegraphics[width=75mm,height=70mm]{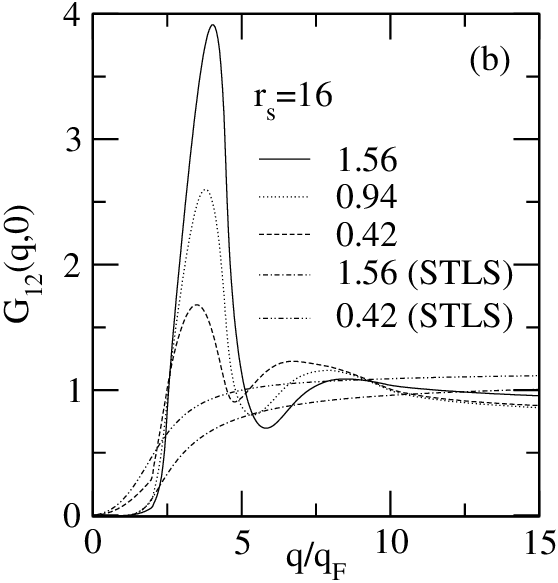}
\caption{\label{fig9}Intralayer (in panel (a)) and interlayer (in panel (b))
static local-field corrections $G_{11}(q,0)$ and $G_{12}(q,0)$ for the
electron-electron bilayer at $r_s=16$ and different layer spacings, according
to qSTLS. Legends indicate the values of $1/\gamma=d/r_sa_0^*$; STLS
local-fields are also shown.}
\end{figure}

The STLS approach again does not give any indication of the WC phase
transition. As in the case of e-h bilayer, this inability of STLS is a
manifestation of its assumption of frequency-independent LFC's. This point is
illustrated in Fig. \ref{fig9} by drawing a comparison between the static
qSTLS LFC's and the STLS LFC's. We notice that $G_{11}(q,0)$ and $G_{12}(q,0)$
both exhibit, in contrast with their STLS counterparts, a pronounced
peaked-structure in the intermediate wavevector region before converging to
their respective limiting values, which are close to the corresponding STLS
limiting values.

\subsection{Dynamic local-fields}
\begin{figure}
\includegraphics[width=75mm,height=70mm]{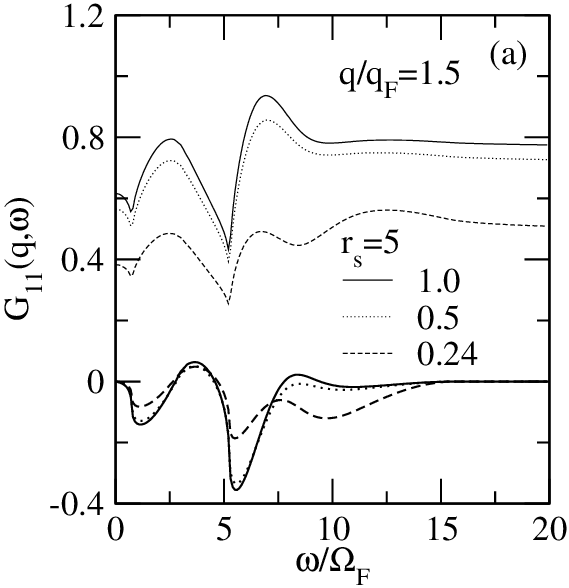}\vspace{5mm}
\includegraphics[width=75mm,height=70mm]{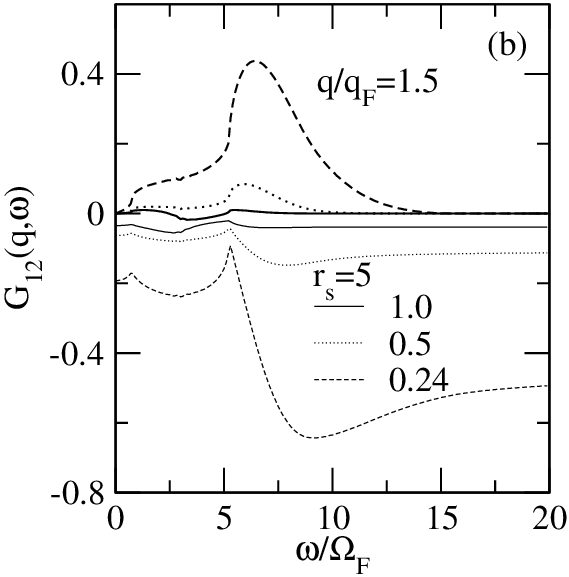}
\caption{\label{fig10}Frequency-dependence of intralayer (in panel (a)) and
interlayer (in panel (b)) local-fields $G_{11}(q,\omega)$ and
$G_{12}(q,\omega)$ for the electron-hole bilayer at $q/q_F=1.5$, $r_s=5$, and
different values of $1/\gamma=d/r_sa_0^*$. Thin and thick lines represent,
respectively, the respective real and imaginary parts; legends indicate the
$1/\gamma$ values and  $\Omega_F$ is the Fermi frequency.}
\end{figure}

\begin{figure}
\includegraphics[width=75mm,height=70mm]{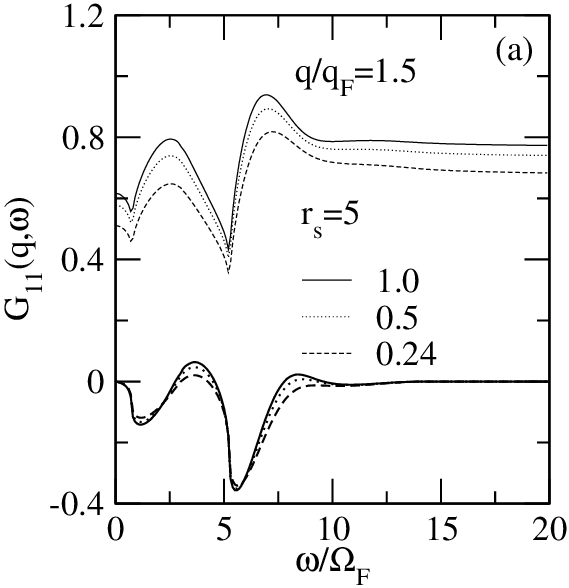}\vspace{5mm}
\includegraphics[width=75mm,height=70mm]{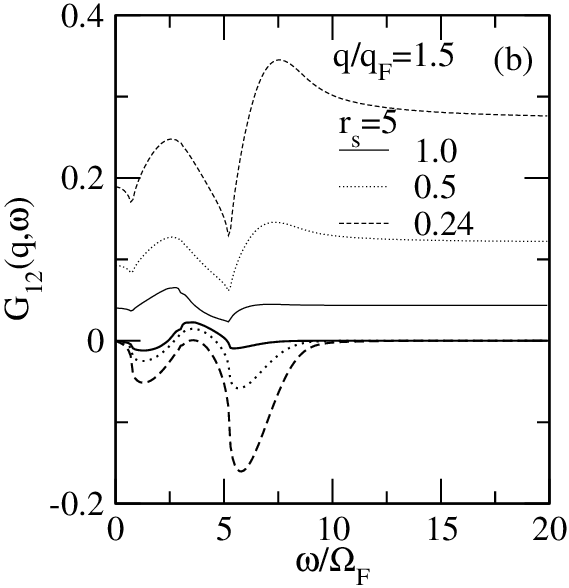}
\caption{\label{fig11}(a)-(b) Dynamic local-fields for the electron-electron
bilayer; Description of curves is exactly the same as in Fig. \ref{fig10}.}
\end{figure}

We have seen above that it is the frequency-dependence of LFC's which brings
in a marked difference in the qSTLS description of correlations as compared to
the STLS one. Therefore, we examine the LFC's for their dependence on
frequency. The results for the e-h and e-e bilayers are reported,
respectively, in Figs. \ref{fig10} and \ref{fig11} at $q/q_F=1.5$, $r_s=5$,
and different $d's$.  The real and imaginary parts of both the intra- and
interlayer LFC's are oscillatory functions of frequency - a feature which is
analogous to that has been found in 3D \cite{r19}, 2D \cite{r11}, and 1D
\cite{r20} electron systems.  In the large frequency limit, both
$G_{11}(q,\omega)$ and $G_{12}(q,\omega)$ approach formally the STLS LFC's.
The LFC's in the e-h and e-e bilayers have qualitatively a similar dependence
on frequency. However, we notice that, apart from the obvious difference of
sign between the interlayer LFC's of the e-h and e-e layers, the attractive
interlayer correlations in the former are relatively stronger in the close
proximity of layers. Similar is the behavior of LFC's at other $q$ and $r_s$
values.

\section{Summary and conclusions}
In summary, we have presented a study of the ground-state behavior of the
symmetric e-e and e-h bilayers by including the effect of dynamic correlations
within the qSTLS theory. We have found that the inclusion of the dynamical
nature of correlations introduces quantitative as well as qualitative
differences in the description of many-body properties as compared to static
mean-field theories of the STLS type.  The qSTLS predictions for the intra-
and interlayer pair-correlation functions and the ground-state energy are
found to be in overall better agreement with the DMC results. The growing
oscillatory trends in the DMC intralayer correlation function with increasing
$r_s$ are accurately reproduced both in terms of amplitude and period for
$r_s\leq 10$ and $1/\gamma \geq 0.5$ - a feature that is missing in the STLS
results. However, the degree of agreement with the DMC data, specifically at
small interparticle separation, becomes somewhat worse with increasing
$\gamma=r_sa_0^*/d$, at a given $r_s$ or $d$, i.e., in the strong coupling
regime.  Another unique and important feature of the qSTLS theory is that, in
both the e-h and e-e bilayers, it exhibits an instability towards a coupled WC
ground state, below a critical density (i.e., for $r_s>r_s^c$) and in the
close proximity of the layers. Moreover, at high density (for $r_s<r_s^c$) it
indicates transition to a CDW ground state. Thus, a crossover from the CDW
instability to the WC instability takes place at $r_s=r_s^c$.  Our prediction
of Wigner crystallization agrees qualitatively with the findings of the DMC
calculations. However, the critical value of $r_s$ (i.e., $r_s^c$) is
underestimated. This discrepancy in the estimate of $r_s^c$ seems to reflect
the misrepresentation of short-range correlations at large $\gamma$ or $r_s$
in the qSTLS theory.  Since the Coulomb coupling among carriers grows with
increasing $\gamma$ or $r_s$ and the exchange correlations are relatively less
important in the strong coupling regime, we believe that the failure of the
qSTLS theory in such a regime arises from its neglect of the dynamics of the
Coulomb correlations, i.e., of the Coulomb correlation hole.  Another
contribution to the failure is possibly due to the neglect of interaction
effects on the momentum distribution function. These issues deserve further
investigation.

Finally, we have also employed the qSTLS theory to examine the
spin-polarization effects in the e-h bilayer. Interestingly enough, a
polarization transition is found to take place from the unpolarized to the
polarized liquid (at $r_s\approx 6$) well before the unpolarized liquid could
actually make transition to the WC ground state. The polarized e-h bilayer too
supports the CDW and WC instabilities, but the crossover density is now
lowered to $r_s\approx 27$.

\paragraph*{Acknowledgments} 

One of us (RKM) gratefully acknowledges the financial
support from the Ministero dell'Universit\'{a} e della Ricerca Scientifica e
Tecnologica.


\begin{thebibliography}{99}
\bibitem{note}In fact, while we were completing the present calculations, we
 have been able to device a manner to handle the insurgence of divergent peaks
 in the response functions, so that we are now able to perform converged
 calculations for values of the physical parameters $r_s$ and $d$ near
 and even inside the instability region.
\end{thebibliography}
\end{document}